\documentclass[twocolumn,showpacs,aps,floatfix]{revtex4}
\usepackage{graphicx}

\def\fun#1#2{\lower3.6pt\vbox{\baselineskip0pt\lineskip.9pt
  \ialign{$\mathsurround=0pt#1\hfil##\hfil$\crcr#2\crcr\sim\crcr}}}
\def\etal{{\it et al.}}
\def\simle{\lower 2pt \hbox {$\buildrel < \over {\scriptstyle \sim }$}}
\def\simge{\lower 2pt \hbox {$\buildrel > \over {\scriptstyle \sim }$}}

\def\lapproxeq{\lower .7ex\hbox{$\;\stackrel{\textstyle <}{\sim}\;$}}

\begin{document}

\title{Correlation of the Highest Energy Cosmic Rays with
 the Supergalactic Plane}

\author {Todor~Stanev}
\address{
Bartol Research Institute, Department of Physics and Astronomy, 
University of Delaware, Newark, DE 19716, USA
}

\vspace*{5truemm}
\begin {abstract}
 We examine the anisotropy of the arrival directions of twenty seven
 ultra high energy cosmic rays detected by the Pierre Auger Collaboration.
 We confirm the anisotropy of the arrival directions of these 
 events and find a  significant correlation with
 the updated definition of the supergalactic plane at distances
 up to 70 Mpc, A Monte Carlo calculation of isotropic source
 distribution suggests a  chance probability for isotropic event
 arrival direction distribution  of 2-6$\times$10$^{-4}$. 
\end{abstract} 

\maketitle
  
  The identification of the sources of the ultrahigh energy cosmic
 rays (UHECR), the highest energy nuclei in the Universe, is 
 one of the most important goals of the high energy astrophysics.
 It is difficult to even imagine how nature can manage to 
 accelerate these particles to energies exceeding by orders
 of magnitude what we can do in man made accelerators. 
 Since these nuclei of energy approaching 10$^{20}$ eV are not
 strongly deflected in the galactic magnetic fields, one expects
 them to come from directions coincidental with their sources
 if extragalactic magnetic fields are not
 very strong. Source identification would not only prove that
 cosmic ray astronomy is possible, but also contain information
 about the magnetic and photon fields in the source direction
 that can not be extracted in any other way. The first step in 
 this direction was made less than a year ago.

  The Auger Collaboration reported a correlation of their
 27 highest energy events (E $>$ 5.7$\times$10$^{19}$ eV (57 EeV)
 with active galactic nuclei (AGN) at redshifts $z$ less than
 0.017~\cite{Auger-Science} from the V\`{e}ron-Cetty and V\`{e}ron
 catalog~\cite{Veron} (V-C). Twenty out of 27 events are within 3.1$^o$
 of individual AGN, while for
 isotropic arrival distribution one expects on average 5 coincidences.
 Five of the non-correlating events come from less than 12$^o$
 galactic latitude which may be understood as larger deflections in
 the galactic magnetic field. Most of the correlating events are
 visibly at relatively small angles from Cen A. It is surprising
 that there are no events coming from the direction of the Virgo
 cluster, that includes a large number of powerful galaxies in
 addition to M87, as stated in Ref.~\cite{Gorbunovetal08}.

 Several papers among them Refs.~\cite{Mosc,MRG,Ghis,Zaw} were
 submitted during the last three months that discussed the type
 of the correlating sources, their luminosity and distance, as
 well as made correlations with different other types of sources
 and attempted to explain the large number of UHECR around Cen A
 and the lack of events close to the Virgo cluster. All sources
 discussed, be them AGN from the V-C or Swift BAT catalogs or
 spiral galaxies, do tend not to be distributed isotropically
 and contribute to the definition of the Supergalactic Plane (SGP) -
 the plane of weight of nearby galaxies~\cite{deV}. Many of the 
 Auger events thus come from directions close to the SGP.
 There are, however, eight UHECR, five of which correlate with AGN,
 at supergalactic latitude higher than 40$^o$. 

 A report on the correlation of the then existing statistics 
 of UHECR of energy above 4$\times$10$^{19}$ eV was published
 in 1995~\cite{Stanevetal95}. The data set consisted mostly
 of events detected by the  Haverah Park detector with the addition
 of events  detected by AGASA, Yakutsk and Volcano Ranch air
 shower arrays.
 The anisotropy of that data set was studied by a comparison of 
 the average and RMS supergalactic latitudes $|b_{SGP}|$ of the
 experimental events to that of an isotropic Monte Carlo sample.
 The significance of the correlation was at the 3$\sigma$ level.
 Later on, when the AGASA detector dominated the
 world's statistics~\cite{Agasa}, the  significance of the
 correlation decreased.

\begin{figure}[thb] 
\centerline{\includegraphics[width=80truemm]{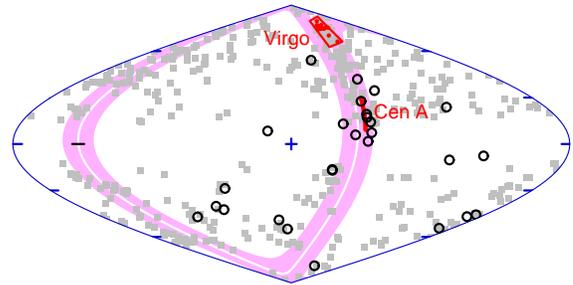}}
\caption{The 27 Auger events are superimposed
 in galactic coordinates in a sinusoidal projection on the 
 supergalactic plane in the definition of Ref.\cite{deV1}.
 The shaded area shows $|b_{SGP}| <$ 10$^o$. The gray squares
 show the positions of V-C AGN that have galactic latitude 
 more than 12$^o$.
\label{fig1}
}
\end{figure}

  After the list of the energies and arrival direction of
 the 27 events was published in a more detailed second 
 paper~\cite{Auger_2} it became possible to study the 
 statistical significance of the possible correlation of
 the Auger events with the Supergalactic Plane.
 The arrival directions of these events are superimposed on the SGP
 in Fig~\ref{fig1}. The extended Cen A radio lobe, almost parallel
 to SGP, and the Virgo cluster (with 11 of its most powerful
 galaxies plotted) are outlined.
  We attempted to calculate the probability for such a correlation
 by a Monte Carlo simulation. We injected 100,000 sets of 27
 events each using the equal field of view areas published
 by the Auger Collaboration~\cite{Auger_2}. 

  Because of the large number of UHECR at large supergalactic 
 latitude we decided a posteriori to count the events that come from 
 directions close to SGP. We chose $|b_{SGP}|$ less than 
 10$^o$ and 15$^o$. Two arbitrary bands with a difference of
 5$^o$ were chosen because the average magnitude of the scattering
 angle in the galactic magnetic field is of that size.
 Since the number of experimentally detected events is small
 we chose to test two bands in order to not be  deceived by
 random coincidences. Nine and 13 events fall in these two bands.
 The probability of random coincidences, based on the isotropically
 distributed Monte Carlo sample, is 0.024 and 0.008
 respectively, corresponding to 2.0 and 2.4 $\sigma$. 
 
\begin{figure}[thb] 
\centerline{\includegraphics[width=80truemm]{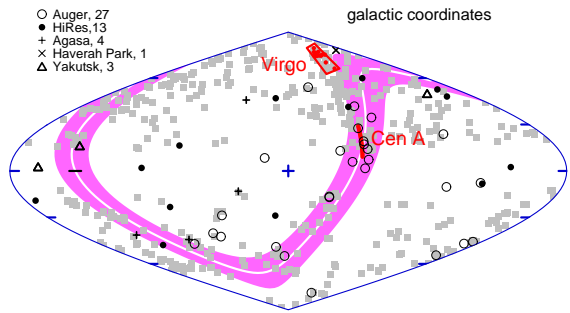}}
\caption{The supergalactic plane corrected as in Ref.~\cite{Lahavetal}
 for distances up to 70 Mpc. The Virgo cluster and Cen A are also shown. 
 The events from Auger are complemented with 13 events from HiRes and
 three small sets of  events from Haverah Park, AGASA and Yakutsk.  
\label{fig2}
}
\end{figure}

 Until now we dealt with the original definition~\cite{deV,deV1}
 of the supergalactic plane by de Vaucouleurs. This definition was
 studied in 2000 by Lahav et al. in terms of redshift~\cite{Lahavetal}
 on the basis of the Optical Redshift Survey (ORS)
 (8457 galaxies, 98\% with redshift). Since ORS had a zone of avoidance
 of $|b| <$ 20$^o$ it was complemented with IRAS galaxies in order
 to describe better the intersection of SGP with the galactic plane.   
 This study introduces correction to the definition of SGP for 
 different distances from 0 to 80 Mpc. The one that is most suitable
 for analysis  of the Auger events is for distances up to 70 Mpc,
 i.e. almost identical to the redshift of 0.017.
 The corrected shape of the SGP is plotted in Fig.~\ref{fig2} the same
 way as the original definition is plotted  in Fig.~\ref{fig1}.
  Although the corrected SGP does not look very different
 from the original one in galactic coordinates the SGP North
 Pole direction is at the original $b_{SGP}$ = 65$^o$.
 The same degree of rotation defines the SGP updated 
 to distances of 80 Mpc.

  One can now count how many UHECR are in the 
 10$^o$ and 15$^o$ band with the new definition of the
 supergalactic plane. The numbers are  13 and 15 events
 respectively. The probability to have so many 
 events drops to 1.6$\times$10$^{-4}$ (6.0$\times$10$^{-4}$) 
 for distances from the supergalactic plane of 10$^o$(15$^o$).
 The correlation of the Auger events with the updated definition
 of the SGP within 70 Mpc is indeed much more significant - the
 probabilities above correspond to 3.6(3.2)$\sigma$. 

  The average supergalactic latitude of the Auger events does not
 change much - from 26.7$^o$ with the classical definition to
 22.5$^o$ with the new one. The RMS supergalactic latitude of
 the experimental events is 32$^o$ compared to 38.9$^o$ for 
 a Monte Carlo sample of 100,000 uniformly distributed event
 samples in the Auger field of view. None of these results has  high
 statistical significance. It is interesting to note that the 
 308 AGN from the V-C catalog within redshift of 0.017 and with 
 galactic latitude more than 12$^o$ do not correlate better with 
 the new definition of SGP - the average supergalactic latitude
 is 24.9$^o$ compared to 22.8$^o$ for the classical definition. 

 Meanwhile we had the report of the HiRes Collaboration~\cite{HiRes1}
 that does not see correlation of their 13 highest energy events
 with AGN. Because of the difference of the energy assignments of
 the two experiment the HiRes Collaboration has roughly adjusted
 the number of events to correspond to the experimental exposure.
 We plot the HiRes events~\cite{HiRes13} with dots in
 Fig~\ref{fig2}. Only two of 13 events are within 3.1$^o$ of an AGN
 from the same catalog. The chance probability for isotropic
 arrival direction distribution quoted in
 Ref.~\cite{HiRes1}  is 82\%. The field of view of HiRes is, of course,
 different from that of Auger, and since it is not presented in
 Ref.~\cite{HiRes1} we cannot estimate the possible correlation 
 with SGP. As could be seen in Fig.~\ref{fig2}, 3(5) events are
 within 10$^o$(15$^o$) of the supergalactic plane. There are,
 however, two pairs of events from Auger and HiRes that are 
 very close to each other - at angular distances of 0.5$^o$ and 4$^o$.
 The probability of these coincidences also cannot be estimated. 

  Since a degree of correlation with the supergalactic plane 
 was found both in the Auger and the old data sets it is worth
 examining the two samples. The most difficult decision is to
 use approximately the same event energy threshold, since both data
 sets were shown to be sensitive to it and the energy assignments of
 all experiments are different. The correlations in the old 
 data set appeared at energy above 4$\times$10$^{19}$ eV. One
 should have in mind that Northern hemisphere detectors 
 look mostly in the direction of the galactic anti-center, thus
 the scattering in the galactic magnetic fields~\cite{stanev97}
 should be smaller.

  We approached the selection of old events using the exposure of the
 old experiments. Agasa~\cite{Agasa,Teshima06}
 has the exposure of 1,400 km$^2$.sr.yr.
 Comparing to the exposure of Auger (9,000 km$^2$.sr.yr) we chose 
 the four highest energy Agasa events. From Yakutsk~\cite{Yakutsk} 
 (850 km$^2$.sr.yr, {\em A.A.~Ivanov, private communication})
 we chose the 3 highest energy events and from Haverah
 Park~\cite{HP} (250 km$^2$.sr.yr, {\em A.A.~Watson, private communication})
 we took one event. The arrival directions of
 these eight events are shown in Fig.~\ref{fig2}.
 Once again, it is difficult to judge the importance of the
 correlation because of the choice of events and the uncertain 
 field of view of the three experiments but 1 event out of 8 is
 within 10$^o$ and 5 are within 15$^o$ of the updated SGP.
 Only 3 events are within 15$^o$ from the classical definition of SGP.
 Although we are not able to estimate the significance of this
 correlation, it obviously becomes somewhat stronger.

 In conclusion, we studied the correlation of the 27 highest energy
 events detected by the Pierre Auger experiment with the supergalactic
 plane. We find 1-2\% significance of the Auger events with the
 classical definition of SGP. The correlation is much stronger
 (2-6$\times$10$^{-4}$ chance probability for isotropic arrival
 direction distribution) of the 27 Auger events
 with the updated definition{~\cite{Lahavetal} of SGP within
 70 Mpc. 

 Five of 13 HiRes events are within 15$^o$ of the updated SGP, while 
 5 out of 8 events from Agasa, Yakutsk and Haverah Park are at similar
 supergalactic latitude. Because of the lack of knowledge about the 
 HiRes field of view and the uncertainty in our choice of events
 from three other experiments we can not estimate correctly the significance 
 of these correlations. It is obvious, though, that the flux of UHECR
 does not appear to be isotropic - a total of 17 (25) out of 48 
 events are within supergalactic latitude of 10$^o$( 15$^o$).
 A degree of correlation with the SGP
 is expected as all suggested sources of UHECR should correlate
 with the local large scale structure. 
  
 It is possible that the five experiments examined here may see 
 events coming from the same sources. There are three possible 
 event pairs to Auger events - two from HiRes and one with Agasa.
 There is also a possible triple that includes two HiRes and one
 Yakutsk events. As far as Auger data is concerned, the largest
 event concentration is around the nearby radio galaxy Centaurus A.
 One can imagine that one third of the 27 Auger events are accelerated
 at this object and are spread along the SGP by low level organized
 extragalactic magnetic fields~\cite{SSE}.

 While performing this study we have not contributed to the 
 identification of the sources of UHECR. We have only studied the
 isotropy of the arrival directions of these participles and have
 confirmed that the arrival directions of the Auger event sample are
 not isotropic using a different analysis procedure. The significance
 of the anisotropy of other event samples can only be studied after
 their fields of view are defined as well as that of the Auger
 Collaboration. 

 The author thanks T.K.~Gaisser, D.~Seckel and A.A.~Watson for very
 helpful discussions on the topic of this communication. This work is 
 supported in part by NASA ATP grant NNG04GK86G.

\end{document}